\documentclass[aps,pra,twocolumn,superscriptaddress,showpacs]{revtex4-1}
\usepackage{amsmath}
\usepackage[dvips,xdvi]{graphicx}
\usepackage{amssymb}
\usepackage{epsfig}

\begin{document}
\title{A simple and efficient all-optical production of spinor condensates}
\author{J. Jiang}
\author{L. Zhao}
\author{M. Webb}
\affiliation{Department of Physics, Oklahoma State University,
Stillwater, OK 74078}
\author{N. Jiang}
\author{H. Yang}
\affiliation{Center for Quantum Information, IIIS, Tsinghua
University, Beijing, China}
\author{Y. Liu}
\email{yingmei.liu@okstate.edu} \affiliation{Department of
Physics, Oklahoma State University, Stillwater, OK 74078}
\date{\today}

\begin{abstract}
We present a simple and optimal experimental scheme for an
all-optical production of a sodium spinor Bose-Einstein condensate
(BEC). With this scheme, we demonstrate that the number of atoms
in a pure BEC can be greatly boosted by a factor of 5 in a simple
setup that includes a single-beam optical trap or a crossed optical
trap. This optimal scheme
avoids technical challenges associated with some all-optical BEC
methods, and can be applied to other optically trappable atomic
species. In addition, we find a good agreement between our theoretical model and data. The upper limit for the efficiency of evaporative cooling in all-optical BEC approaches is also discussed.
\end{abstract}


\maketitle In the last two decades, many techniques have been
developed to reliably generate a BEC of more than $10^4$ atoms.
Almost every one of these techniques requires evaporative cooling
in a trapping potential, including a magnetic trap, an optical
dipole trap (ODT), or a combined magnetic and optical
potential~\cite{DavisBEC, Anderson, Bradley, ketterle, hybrid}.
Among these techniques, all-optical methods have been proven to be
versatile and popularly applied in producing quantum
degenerate gases of both bosonic~\cite{Barrett, Dumke, Weber, Adams,
Arnold, Kinoshita, Clement, Takasu, Olson} and
fermionic~\cite{Granade} species. ODTs have tight confinement
which allows for fast evaporation with a duty cycle of a few
seconds~\cite{Barrett}. Unlike magnetic potentials that only
trap atoms in the weak-field seeking spin state, an ODT can
confine all spin components. This is crucial for creating vector
(spinor) BECs with a spin degree of freedom~\cite{Stenger}. ODTs
can also be applied to a wider variety of atomic species (e.g.,
Ytterbium, alkaline earth metals, and Cesium) which cannot
be feasibly condensed in a magnetic trap~\cite{Weber, Takasu}. In addition,
optical trapping does not require magnetic coils around trapped
atoms, which not only provides better optical access but also
yields less residual magnetic fields. The simplicity and
versatility of ODTs widens the accessibility of BEC research on
many-body physics, precision measurements, and quantum information
science~\cite{spinor}.
\begin{figure}[t]
\includegraphics[width=85mm]{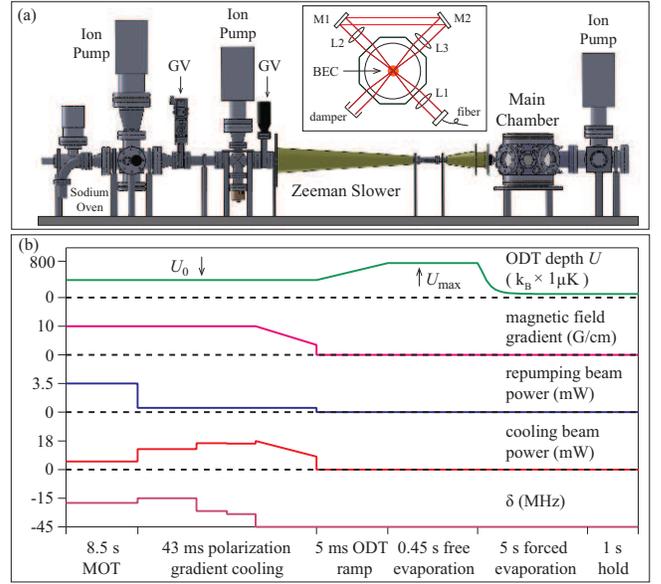}
\caption{(color online) (a). Schematic of our apparatus. GV stands
for a pneumatic gate valve. Inset: schematic of the crossed ODT
setup around the main chamber. L$_1$, L$_2$, and L$_3$ are convex
lenses. M$_1$ and M$_2$ are mirrors. (b). Experimental sequence of
creating sodium BECs with the all-optical approach (see text).
Each MOT cooling beam is detuned by $\delta$ from the cycling
transition. All axes are not to scale.} \label{design}
\end{figure}

Forced evaporation in an ODT can be performed by simply reducing its
trap depth $U$ (e.g., lowering the trapping laser power). In this process, collision rates decrease with $U$, which leads to slow rethermalization and eventually stagnation in evaporative cooling. Several methods have been reported to overcome this difficulty, including tilting an ODT with a magnetic field gradient~\cite{Hung}, using a misaligned crossed
ODT~\cite{Clement, Olson}, compressing an ODT with a mobile lens~\cite{Kinoshita}, or applying multiple ODTs for staged evaporation~\cite{Arnold,Weber}. In this paper, however, we show that
these methods may not be necessary. Good agreements between our
model and experimental data enable us to develop an optimal ODT
ramp and evaporation sequence for an all-optical production of
sodium BECs. With this sequence, we find that the number of atoms
in a pure BEC is greatly boosted and evaporation efficiency $\gamma=-d(\ln D)/d(\ln N)$ can be 3.5 in a simple setup that includes a single-beam ODT or a crossed ODT. Here $D$ is the phase space density and $N$ is the atom number. We also show an upper limit for $\gamma$ at a given
truncation parameter $\eta=U/\rm{k_B T}$, and demonstrate that a
constant $\eta$ does not yield more efficient evaporative cooling.
Here $\rm{T}$ is the atom temperature and $\rm{k_B}$ is the Boltzmann
constant. This optimal experimental scheme allows us to avoid technical
challenges associated with some all-optical BEC approaches.

Figure~\ref{design}(a) shows a schematic of our apparatus, which
is divided via differential pumping tubes and gate valves into an
atomic oven chamber, an intermediate chamber, and a main chamber
where a magneto-optical trap (MOT) is located. The intermediate
chamber allows us to refill alkali metals and get ultra-high
vacuum (UHV) pressures back to the 10$^{-12}$ Torr range within 24
hours. A typical experimental sequence of our all-optical BEC
approach is shown in Fig.~\ref{design}(b). Hot atoms are slowed
down by a spin-flip Zeeman slower as elaborated in
Ref.~\cite{slower}. The slowed atoms are collected in a standard
MOT which is constructed with six cooling beams and a pair of
24-turn anti-Helmholtz coils. Each MOT cooling beam is detuned by
$\delta=-20$~MHz from the cycling transition, has a power of 6~mW,
and combines with one 3.5~mW MOT repumping beam in a single-mode
fiber. After 8.5~s of MOT loading, a three-step polarization
gradient cooling process efficiently cools $3\times 10^8$
atoms to 40 $\mu$K~\cite{PGC}. To depump atoms into the F$=$1
hyperfine states, the repumping beams are extinguished 1~ms before
cooling beams and MOT coils are turned off.

A crossed ODT consists of two far-detuned beams which originate from an infrared (IR) laser at 1064~nm and have a
$1/\rm{e}^2$ waist of 33~$\mu$m at their intersection point, as
shown in Fig.~\ref{design}(a). A single-mode
polarization-maintaining fiber is used to polish the beam mode,
and to minimize pointing fluctuations due to imperfections of the
IR laser and thermal contractions of an acoustic-optical
modulator. As a result, atoms which are transferred from the MOT
into the tightly-focused crossed ODT demonstrate a long lifetime
of 8~s and a large collision rate. These are essential for
all-optical BEC approaches.

\begin{figure}[t]
\includegraphics[width=85mm]{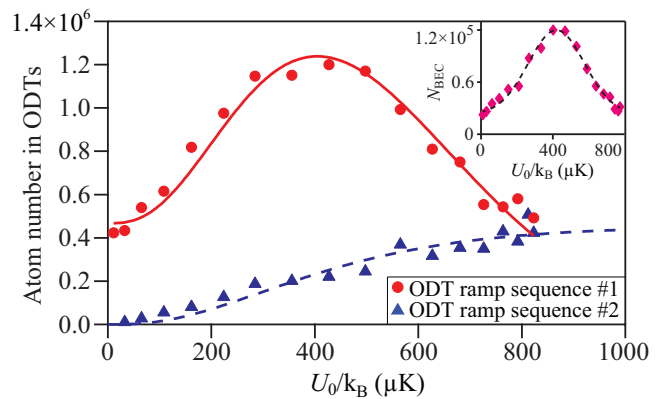}
\caption{(color online) The number of atoms transferred to the
crossed ODT as a function of $U_0$ from two ODT ramp sequences.
Our optimal ramp sequence is the best scenario of the first ODT
ramp sequence with $U_0\simeq U_{\rm{max}}/2$. The solid (red)
line and the dashed (blue) line are fits based on $N_{\rm ramp1}$
and $N_{\rm ramp2}$, respectively (see text). Inset: the number of
atoms in a BEC as a function of $U_0$, when the first ODT ramp
sequence and a same evaporation curve are applied. The dashed
(black) line is a Gaussian fit to data.} \label{fig:AC}
\end{figure}

We compare the efficiency of transferring laser-cooled atoms to
the crossed ODT, when different experimental sequences are applied
to ramp up the ODT's trap depth. Our data in Fig.~\ref{fig:AC}
clearly shows that an optimal ODT ramp sequence can increase the number of atoms loaded into the crossed ODT by a
factor of 2.5. This optimal ramp sequence is the best scenario of
our first ODT ramp sequence. As shown in Fig.~\ref{design}(b), the ODT in the first ODT ramp sequence is kept at a small trap
depth $U_0$ during the entire laser cooling process and then
linearly ramped to $U_{\rm{max}}$ in $t_{\rm{ramp}}=5$~ms. $U\rm{_{max} \approx k_B\times 800~\mu K}$
is the maximal trap depth used in this work and $0\leq~U_0\leq
U_{\rm max}$. This optimal ramp sequence is a trade-off between
two competing effects induced by the enormous ODT potential.
If the ODT beams do not interact with the MOT, a larger $U$
enables more atoms to be transferred to the ODT. The number of
atoms loaded in the ODT is $N_{\rm ramp2} \sim \int_0^{U} \rho
(\epsilon) f (\epsilon) d \epsilon$, where $\rho (\epsilon)$ and
$f(\epsilon)$ are the density of states and occupation number at
energy $\epsilon$, respectively. This is confirmed by our data
(closed blue triangles in Fig.~\ref{fig:AC}) taken with the second
ODT ramp sequence, in which the ODT is linearly ramped in 5~ms
from 0 to $U_0$ immediately after MOT beams are switched off. On
the other hand, if intense ODT beams and MOT beams are turned on
simultaneously, atoms experience non-negligible AC Stark shifts in
regions where the ODT beams and the MOT overlap. As a result, the
MOT's cooling capability is impaired in the MOT and ODT
overlapping regions, and the number of atoms loaded into the ODT
decreases when the ODT becomes too deep. $N$ is thus not a
monotonic function of $U$, which is the case for our first ODT ramp sequence. The number of atoms loaded into the ODT in the first ramp sequence may be expressed as $N_{\rm ramp1}\sim A \xi_{\rm AC}(U_0) \int_0^{U_0} \rho (\epsilon) f(\epsilon)
d \epsilon + \int_{U_0}^{U_{\rm max}} \rho(\epsilon) f(\epsilon)
d\epsilon$. Here $\xi_{\rm AC}(U_0) = \exp(-(\delta_{\rm
AC}(U_0))^2/\omega_0^2)$ is a correction factor due to the induced
AC Stark shift $\delta_{\rm AC}(U_0)$, while $A$ and $\omega_0$
are fitting parameters. Our data (closed red circles in
Fig.~\ref{fig:AC}) collected with the first ODT ramp sequence can
be well fit by this model. The fit value of $\omega_0$ is
$1.2\Gamma$, where $\Gamma/2 \pi =9.7~$MHz is the natural
linewidth of sodium. The number of atoms in the ODT reaches its
peak when the optimal ramp sequence with $U_0\simeq
U_{\rm{max}}/2$ is applied. Compared to the second ODT ramp
sequence, the optimal ramp sequence allows us to use ODT beams
with smaller waists while loading the same amount of laser-cooled
atoms to the ODT. The resulted high initial atomic density and
high collision rates from the optimal ramp sequence enable
very efficient evaporative cooling. This greatly boosts the number
of atoms in a BEC by a factor of 5 and a good evaporation efficiency $\gamma=3.5$ is achieved, as shown in the inset of
Fig.~\ref{fig:AC}.

\begin{figure}[t]
\includegraphics[width=85mm]{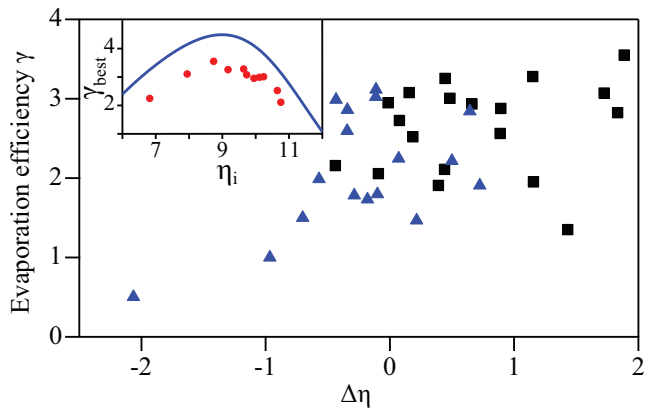}
\caption{(color online) Evaporation efficiency $\rm{\gamma}$ in 36
different evaporation processes as a function of
$\rm{\Delta\eta}$. Closed (black) squares are data taken with the
forced evaporation time longer than 1~s. Inset:
$\rm{\gamma_{best}}$ as a function of $\rm{\eta_i}$ extracted from
the main figure. The solid line sets an upper limit for $\gamma$
based on Eq.~(\ref{eqn:r}) by assuming $k_1=k_{\rm s}=0$ (see text). }
\label{fig:psd}
\end{figure}

To optimize $\gamma$, it is necessary to understand the time evolution of the system energy $E$ and the atom number $N$ during an evaporation process. Similar to Ref.~\cite{Ohara, Luiten, Yan,
Olson}, we use $\rm{\kappa k_BT\approx(\eta-5)/(\eta-4)k_BT}$ to
represent the average kinetic energy taken by an atom when it is
removed from the ODT, and assume the mean kinetic energy and
mean potential energy to be $E/2$ when $\eta$ is large. The time
evolution of $E$ and $N$ is thus given by,
\begin{align} \label{eqn:e}
\dot E=&-\frac{2 (\eta-4) {\rm e}^{-\eta} N}{\tau_2} (U+\kappa
{\rm k_BT})+\frac{\dot{U}}{U}\frac{E}{2}+\dot{E}|_{\rm{loss}}~, \nonumber \\
\dot N=&-2 (\eta-4) {\rm e}^{-\eta} N/\tau_2+\dot{N}|_{\rm{loss}}~,
\end{align}
where $\tau_2$ is the time constant of the two-body elastic collision.
In Eq.~\ref{eqn:e}, $\dot{E}|_{\rm{loss}}$ and $\dot{N}|_{\rm{loss}}$ are due to various inelastic loss mechanisms and may be expressed as,
\begin{align} \label{eloss}
\dot{E}|_{\rm{loss}}=&k_{\rm s}N-k_1N(3{\rm k_B T}) - k_3 n^2 N (2 \rm{k_B
T})~, \nonumber \\
\dot{N}|_{\rm{loss}}=& - k_1N - k_3 n^2 N~,
\end{align}
where $k_1$ and $k_3$ are one-body and three-body loss rates,
respectively. $k_{\rm s}$ represents heating introduced by ODT beams via
a number of different mechanisms, such as pointing fluctuations of
the ODT beams, bad laser beam mode, and spontaneous light
scattering. The term $\rm{2k_B T}$ in Eq.~(\ref{eloss}) accounts
for the fact that atoms in the ODT's center have higher density
and thus are more affected by the three-body inelastic
loss~\cite{Clement}.
\begin{figure}[t]
\includegraphics[width=85mm]{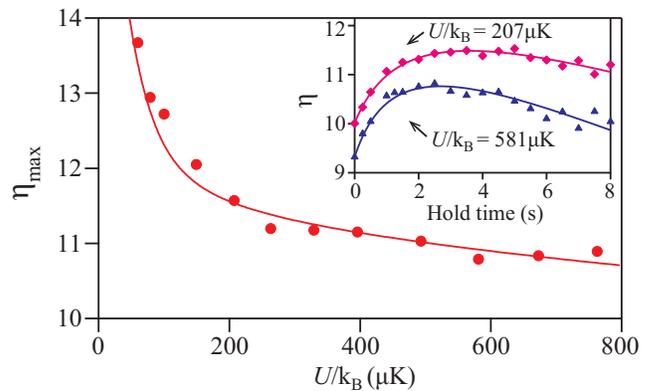}
\caption{(color online) $\rm{\eta_{max}}$ as a function of the ODT
depth $U$, when atoms are held at a fixed $U$ for 8~s. The solid line
is a fit based on Eqs.~(\ref{eqn:e}-\ref{eloss}) (see text). Inset: the time evolution
of $\eta$ at two typical ODT depths. Solid lines are fits based on
the same model applied in the main figure (see text).}
\label{fig:holding}
\end{figure}

\begin{figure*}
\includegraphics[width=180mm]{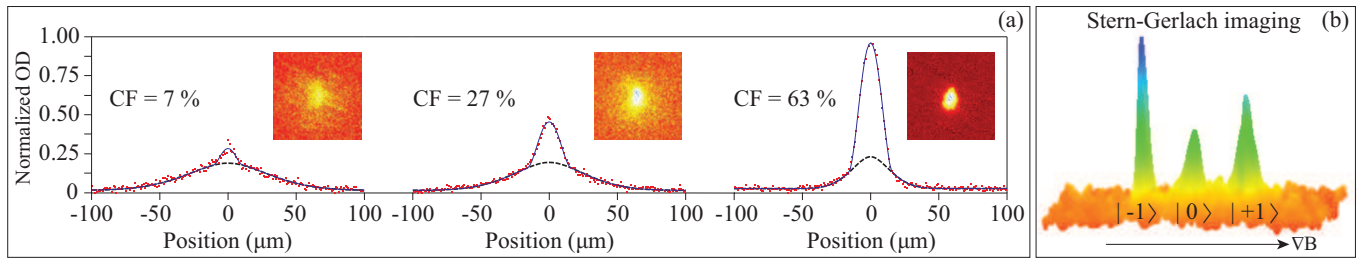}
\caption{(color online)(a). Absorption images taken after
interrupting an optimized evaporation curve at various $U$ followed
by a 10~ms time-of-flight (see text). OD stands for the optical
density. Dashed (black) lines and solid (blue) lines are fits to
the column densities based on a Gaussian distribution and a
bimodal distribution, respectively.
$\rm{CF=\widetilde{n}_{c}/(\widetilde{n}_{th}+\widetilde{n}_{c})}$,
where $\rm{\widetilde{n}_{th}}$ and $\rm{\widetilde{n}_{c}}$ are
column densities for the thermal cloud and the condensate,
respectively. (b). Three spin components of a F=1 spinor BEC are
spatially separated in a 3D Stern-Gerlach absorption image.}
\label{fig:transition}
\end{figure*}
In our apparatus with the UHV pressure in the 10$^{-12}$ Torr range,
background collisions are negligible. Since the ODT beams are
delivered via a single-mode polarization maintaining fiber,
heating induced by the ODT beams is minimized. $k_1$ and $k_{\rm s}$ are
thus very small. If we ignore $k_1$ and $k_{\rm s}$, Eq.~(\ref{eqn:e})
can be simplified to
\begin{equation}  \label{eqn:eeff}
\dot{E}=\dot{N} \eta_{\rm eff} {\rm k_B T} +
\frac{\dot{U}}{U}\frac{E}{2}~,
\end{equation}
where $\eta_{\rm eff}=\eta+\kappa-R(\eta+\kappa-2)$. We define $R=(\dot{N}|_{\rm
loss})/\dot{N}=1/(1+2(\eta-4)e^{-\eta}R_{\rm gTb})$ to represent
the portion of atom losses due to inelastic collisions, where
$R_{\rm{gTb}}$ is the ratio of inelastic collision time constant
to $\rm{\tau_2}$. From solving the above equations, $\gamma$ may be expressed as,
\begin{equation}  \label{eqn:r}
\gamma=\eta_{\rm eff}-4=\eta+\kappa-R(\eta+\kappa-2)-4~,
\end{equation}

The value of $\eta$ in many publications on optical productions of
BECs was held constant with $\rm{\Delta\eta}=0$~\cite{Hung,
Barrett, Kinoshita, Clement, Dumke, Granade, Olson}. Our data in
Fig.~\ref{fig:psd}, however, shows that a constant $\eta$ does not
lead to better evaporation or a larger $\gamma$. The values of
$\gamma$ in this figure are extracted from 36 evaporation
processes in which the forced evaporation speed and the hold time at $U_{\rm max}$ are changed independently, although they all start with the same initial number of
cold atoms in the crossed ODT. $\rm{\Delta\eta}=\rm{\eta_f-\eta_i}$ is the change of $\eta$ during forced evaporation, where $\rm{\eta_i}$ and $\rm{\eta_f}$ are the values of $\eta$ at $U_{\rm max}$ (i.e., the beginning of forced evaporation), and at $U_{\rm f}$, respectively. In order to avoid overestimating $\gamma$ due to the Bose enhancement near the BEC transition temperature, we choose $U_{\rm f}={\rm k_B}\times30~\mu $K where no BEC appears. We
find that $\rm{\Delta\eta}$ tends to be a non-negative value when
the forced evaporation time is longer than 1~s (closed black
squares in Fig.~\ref{fig:psd}), which is a good indication of
sufficient rethermalization. We also find that $\gamma$ is too
small to yield a BEC when $\rm{\Delta\eta}<-2.5$.

We compare the evaporation efficiency at different values of $\rm \eta_i$, as shown in the inset of Fig.~\ref{fig:psd}. $\rm \gamma_{best}$ (the best achieved value of
$\gamma$ at a given $\eta_i$ in our system) does not show a strong dependence on $\eta_i$ if $8<\eta_i<10$, while
$\rm{\gamma_{best}}$ sharply diminishes when $\eta_i$ becomes too
large or too small. In the inset of
Fig.~\ref{fig:psd}, the similar relationship between $\gamma$ and
$\eta_i$ is also predicted by the solid (blue) line, which is a result
based on Eq.~(\ref{eqn:r}) by ignoring $k_1$ and $k_{\rm s}$ and
by applying a non-zero $R$ (i.e.
$R_{\rm{gTb}}=4000$~\cite{ketterle}). All of our data lie below
the solid line in this figure, which may indicate that $k_1$ and $k_{\rm s}$ are larger than 0 and cannot be
ignored. Therefore, based on Fig.~\ref{fig:psd}, we need to choose a value between 8 and 10 for $\eta_i$ and keep $\rm{\Delta\eta}$ larger than -0.5 in order to optimize evaporation efficiency $\gamma$.

The maximum achievable value for $\rm \eta_i$ appears to be 10.8, as shown in the inset of Fig.~\ref{fig:psd}. To understand this, we monitor the time
evolution of $\eta$ and find that $\eta$ has a maximal value
($\rm{\eta_{max}}$) at a given ODT depth $U$. The value of
$\rm{\eta_{max}}$ decreases exponentially with $U$ and
$\rm{\eta_{max}}$ at $U_{\rm{max}}$ is 10.8, which agrees well
with our theoretical prediction (solid red line in
Fig.~\ref{fig:holding}). Therefore, if one wishes to keep $\eta$ unchanged during forced evaporation, $\eta$ must be limited to 10.8 even though
$\rm{\eta_{max}}$ can be much higher at low ODT depths (e.g.,
$\rm{\eta_{max}}>13$ for $U/\rm{k_B<100~\mu K}$). This may be one reason why a constant $\eta$ does not yield more efficient evaporative cooling. We also find that the time evolution of $\eta$ at every $U$ discussed in this
paper can be well fit with our model. Two typical fitting curves
are shown in the inset of Fig.~\ref{fig:holding}.

A pure F=1 BEC of $1.2\times10^5$ sodium atoms at 50~nK is
created from a 0.45~s free evaporation at $U_{\rm{max}}$ followed
by a 5~s forced evaporation in which $U$ is exponentially reduced. This evaporation curve provides two
important parameters for efficient evaporative cooling: $\eta_i$ is
between 8 and 10, and the
forced evaporation time is long enough for sufficient
rethermalization but short enough to avoid excessive atom losses.
Three time-of-flight absorption images in
Fig.~\ref{fig:transition}(a) show a typical change of the
condensate fraction (CF) after interrupting the evaporation curve
at various $U$. When an external magnetic field gradient is applied, three
spin components in a F=1 BEC spatially separate, as shown in
Fig.~\ref{fig:transition}(b). We also apply the above all-optical
approach to evaporate atoms in a single-beam ODT. A similar result can also be achieved in the single-beam ODT as long as its
$1/\rm{e}^2$ beam waist is smaller than 16~$\mu$m to provide a
high enough collision rate.

In conclusion, we have presented an optimal experimental scheme
for an all-optical production of sodium spinor BECs. With this
scheme, the number of atoms in a pure BEC is greatly boosted by
a factor of 5 and $\gamma=3.5$ is achieved in a simple setup that
includes a single-beam ODT or a crossed ODT. We have showed an
upper limit for $\gamma$ at a given $\eta$, demonstrated that a
constant $\eta$ could not yield a larger $\gamma$, and discussed
good agreements between our model and experimental data. This optimal scheme avoids technical challenges associated with some all-optical BEC approaches, and can be applied to other optically
trappable atomic species and molecules~\cite{note}.

We thank the Army Research Office, Oklahoma Center for the
Advancement of Science \& Technology, and Oak Ridge Associated
Universities for financial support. M. Webb thanks the Niblack
Research Scholar program. N. Jiang and H. Yang thank the National
Basic Research Program of China.


\begin{thebibliography}{25}
\expandafter\ifx\csname natexlab\endcsname\relax\def\natexlab#1{#1}\fi
\expandafter\ifx\csname bibnamefont\endcsname\relax
  \def\bibnamefont#1{#1}\fi
\expandafter\ifx\csname bibfnamefont\endcsname\relax
  \def\bibfnamefont#1{#1}\fi
\expandafter\ifx\csname citenamefont\endcsname\relax
  \def\citenamefont#1{#1}\fi
\expandafter\ifx\csname url\endcsname\relax
  \def\url#1{\texttt{#1}}\fi
\expandafter\ifx\csname urlprefix\endcsname\relax\def\urlprefix{URL }\fi
\providecommand{\bibinfo}[2]{#2}
\providecommand{\eprint}[2][]{\url{#2}}
\bibitem{DavisBEC} K. B. Davis, M. -O. Mewes, M. R. Andrews, N. J. van Druten, D. S. Durfee, D. M. Kurn, and W. Ketterle, \bibinfo{journal}{Phys. Rev. Lett.}
  \textbf{\bibinfo{volume}{75}}, \bibinfo{pages}{3969} (\bibinfo{year}{1995}).

\bibitem{Anderson} M. H. Anderson, J. R. Ensher, M. R. Matthews, C. E. Wieman, and E. A. Cornell, \bibinfo{journal}{Science}
  \textbf{\bibinfo{volume}{269}}, \bibinfo{pages}{198} (\bibinfo{year}{1995}).

\bibitem{Bradley} C. C. Bradley, C. A. Sackett, J. J. Tollett, and R. G. Hulet, \bibinfo{journal}{Phys. Rev. Lett.} \textbf{\bibinfo{volume}{75}}, \bibinfo{pages}{1687} (\bibinfo{year}{1995}).

\bibitem{ketterle} W. Ketterle and N. J. van Druten, \bibinfo{journal}{Advances In Atomic, Molecular, and Optical Physics} \textbf{\bibinfo{volume}{37}}, \bibinfo{pages}{181} (\bibinfo{year}{1996}).

\bibitem{hybrid} Y.-J. Lin, A. R. Perry, R. L. Compton, I. B. Spielman, and J. V. Porto, \bibinfo{journal}{Phys. Rev. A} \textbf{\bibinfo{volume}{79}}, \bibinfo{pages}{063631} (\bibinfo{year}{2009}).

\bibitem{Barrett} M. D. Barrett, J. A. Sauer, and M. S. Chapman, \bibinfo{journal}{Phys. Rev. Lett.} \textbf{\bibinfo{volume}{87}}, \bibinfo{pages}{010404} (\bibinfo{year}{2001}).

\bibitem{Dumke} R. Dumke, M. Johanning, E. Gomez, J. D. Weinstein, K. M. Jones, and P. D. Lett, \bibinfo{journal}{New Journal of Physics} \textbf{\bibinfo{volume}{8}}, \bibinfo{pages}{64} (\bibinfo{year}{2006}).

\bibitem{Weber} T. Weber, J. Herbig, M. Mark, H.-C. N\"{a}gerl, and R. Grimm, \bibinfo{journal}{Science} \textbf{\bibinfo{volume}{299}}, \bibinfo{pages}{232} (\bibinfo{year}{2003}).

\bibitem{Adams} C. S. Adams, H. J. Lee, N. Davidson, M. Kasevich, and S. Chu, \bibinfo{journal}{Phys. Rev. Lett.} \textbf{\bibinfo{volume}{74}}, \bibinfo{pages}{3577} (\bibinfo{year}{1995}).

\bibitem{Arnold} K. J. Arnold and M. D. Barrett, \bibinfo{journal}{Optics Communications} \textbf{\bibinfo{volume}{284}}, \bibinfo{pages}{3288} (\bibinfo{year}{2011}).

\bibitem{Kinoshita} T. Kinoshita, T. Wenger, and D. S. Weiss, \bibinfo{journal}{Phys. Rev. A} \textbf{\bibinfo{volume}{71}}, \bibinfo{pages}{011602} (\bibinfo{year}{2005}).

\bibitem{Clement} J.-F. Cl\'ement, J.-P. Brantut, M. Robert-de-Saint-Vincent, R. A. Nyman, A. Aspect, T. Bourdel, and P. Bouyer, \bibinfo{journal}{Phys. Rev. A} \textbf{\bibinfo{volume}{79}}, \bibinfo{pages}{061406} (\bibinfo{year}{2009}).

\bibitem{Takasu} Y. Takasu, K. Maki, K. Komori, T. Takano, K. Honda, M. Kumakura, T. Yabuzaki, and Y. Takahashi, \bibinfo{journal}{Phys. Rev. Lett.} \textbf{\bibinfo{volume}{91}}, \bibinfo{pages}{040404} (\bibinfo{year}{2003}).

\bibitem{Olson} A. J. Olson, R. J. Niffenegger, and Y. P. Chen, \bibinfo{journal}{Phys. Rev. A} \textbf{\bibinfo{volume}{87}}, \bibinfo{pages}{053613} (\bibinfo{year}{2013}).

\bibitem{Granade} S. R. Granade, M. E. Gehm, K. M. O'Hara, and J. E. Thomas, \bibinfo{journal}{Phys. Rev. Lett.} \textbf{\bibinfo{volume}{88}}, \bibinfo{pages}{120405} (\bibinfo{year}{2002}).

\bibitem{Stenger} J. Stenger, S. Inouye, D. M. Stamper-Kurn, H.-J. Miesner, A. P. Chikkatur, and W. Ketterle, \bibinfo{journal}{Nature} \textbf{\bibinfo{volume}{396}}, \bibinfo{pages}{345} (\bibinfo{year}{1998}).

\bibitem{Hung} C.-L. Hung, X. Zhang, N. Gemelke, and C. Chin, \bibinfo{journal}{Phys. Rev. A} \textbf{\bibinfo{volume}{78}}, \bibinfo{pages}{011604} (\bibinfo{year}{2008}).

\bibitem{spinor} D. M. Stamper-Kurn, M. Ueda, \bibinfo{volume}{arXiv:1205.1888} (\bibinfo{year}{2012}).

\bibitem{slower} L. Zhao, J. Jiang, J. Austin, M. Webb, Y. Pu, and Y. Liu, \bibinfo{journal}{to be published} (\bibinfo{year}{2013}).

\bibitem{PGC} The first polarization
gradient cooling step compresses the
MOT for 20~ms by increasing the power of each cooling beam to
12~mW while changing $\delta$ to $-15$~MHz. In this step, the
power of each MOT repumping beam is also drastically reduced to 45
$\mu$W. Then during a 5~ms pre-molasses step, every cooling beam
is further red detuned in addition to its power being increased to
16~mW. This is followed by a 18~ms optical molasses, in which a
cooling beam is detuned to $\delta=-45$~MHz and its power linearly
drops to 8~mW. The magnetic field gradient is also reduced to
3~G/cm over the 18~ms.

\bibitem{Ohara} K. M. O'Hara, M. E. Gehm, S. R. Granade, and J. E. Thomas, \bibinfo{journal}{Phys. Rev. A} \textbf{\bibinfo{volume}{64}}, \bibinfo{pages}{051403} (\bibinfo{year}{2001}).

\bibitem{Luiten} O. J. Luiten, M. W. Reynolds, and J. T. M. Walraven, \bibinfo{journal}{Phys. Rev. A} \textbf{\bibinfo{volume}{53}}, \bibinfo{pages}{381} (\bibinfo{year}{1996}).

\bibitem{Yan} M. Yan, R. Chakraborty, A. Mazurenko, P. G. Mickelson, Y. N. Martinez de Escobar, B. J. DeSalvo, and T. C. Killian, \bibinfo{journal}{Phys. Rev. A} \textbf{\bibinfo{volume}{83}}, \bibinfo{pages}{032705} (\bibinfo{year}{2011}).

\bibitem{note} Upon completion of this work, we have recently become aware of
Ref~\cite{Jacob} and their ODT ramp sequence which linearly ramps
ODTs from $U_{\rm max}/3$ to $U_{\rm max}$ in $t_{\rm ramp}$=2~s. We
find that the number of atoms in a pure BEC exponentially
decreases with $t_{\rm ramp}$ when $t_{\rm ramp}>0.01~$s in our system. The optimal sequence explained in our paper yields a pure BEC of
at least twice the number of atoms than that from a sequence with $t_{\rm ramp}$=2~s for our apparatus.

\bibitem{Jacob} D. Jacob, E. Mimoun, L. D. Sarlo, M. Weitz, J. Dalibard, and F. Gerbier, \bibinfo{journal}{New Journal of Physics} \textbf{\bibinfo{volume}{13}}, \bibinfo{pages}{065022} (\bibinfo{year}{2011}).
\end{thebibliography}
\end{document}